\documentstyle[preprint,aps]{revtex}
\tighten
\begin{document}
\draft
\title{Stuffed Black Holes}
\author{$^{(1)}$A. Arbona, $^{(1)}$C. Bona, $^{(1)}$J. Carot,$^{(1)}$ L. Mas, 
$^{(2,3)}$J. Mass\'o and $^{(1)}$J. Stela}
\address{
$^{(1)}$Departament de Fisica, Universitat de les Illes Balears.\\
	07071 Palma de Mallorca, Spain.
$^{(2)}$Max-Planck-Institut f{\"u}r Gravitationsphysik. \\
	Schlaatzweg 1, 14473 Potsdam, Germany. \\
$^{(3)}$National Center for Supercomputer Applications, Beckman Institute.\\
	405 N. Matthews Ave., Urbana, IL 61801\\
}

\maketitle
\begin{abstract}
Initial data corresponding to spacetimes containing black holes are
considered in the time symmetric case. The solutions are obtained by matching
across the apparent horizon different, conformally flat, spatial metrics. 
The exterior metric is the vacuum solution obtained by the 
well known conformal imaging method. The interior metric for every black hole 
is regular everywhere and corresponds to a positive 
energy density. The resulting matched solutions cover then the whole
initial (Cauchy) hypersurface, without any singularity, and can be useful for
numerical applications. The simpler cases of one black hole (Schwarzschild
data) or two identical black holes (Misner data) are explicitly solved.
A procedure for extending this construction to the multiple black hole 
case is also given, and it is shown to work for all time symmetric vacuum 
solutions obtained by the conformal imaging method. The numerical evolution of
one such 'stuffed' black hole is compared with that of a pure vacuum or 'plain'
black hole in the spherically symmetric case.
\end{abstract}

\narrowtext
\section{INTRODUCTION AND OVERVIEW}

Black holes are the most elementary objects in General Relativity. Allowing for
the well known 'no hair' theorems, they can be characterized by their mass,
charge and spin, like elementary particles in classical or quantum mechanics.
This is because event horizons act as a one way membrane which separates the
black hole interior from the exterior: there is no causal effect that can
propagate from the inner to the outer region. Therefore, one does not need to 
know all the details about physical processes taking place at the interior in 
order to describe the overall evolution of the black hole. This also means that
black hole metrics that differ only inside the holes can lead to the same
exterior spacetime. The idea of this paper is to replace the singular vacuum
solution for a black hole interior by a regular one corresponding to a
non-vanishing energy density, but keeping the same exterior metric. The term
'stuffed' black hole is a good description of the resulting solution.  

The 'no hair' property has been very useful in Numerical Relativity to avoid
the interior black hole singularity appearing in the initial data.
The standard practice is to excise the inner region from the computational 
domain, so that one can safely evolve the exterior (usually vacuum) region. 
This has been done for a single black hole (Schwarzschild case) by setting an 
internal boundary on the computational domain at the initial position $r_0$ of 
the apparent horizon. The same thing has been done for a system consisting of 
two non rotating black holes, starting the evolution from the Misner 
data~\cite{Misner,Smarr}, or even for an arbitrary number of them, by using 
the 'conformal imaging formalism'~\cite{Lindquist} to construct conformally 
flat initial data. We shall refer to all these pure vacuum solutions as
'plain' black hole metrics to distinguish them from the 'stuffed' ones we are
presenting here.

In this paper we propose to take advantage of the 'no hair' property in a
different way. We will keep the same exterior metric of the 'plain' black hole
case, obtained by the conformal imaging formalism. But we will match this 
exterior metric to a regular conformally flat interior corresponding to a 
positive energy density. 
The exterior solution is not affected because the matching surfaces
coincide in every case with the initial position $r_0$ of the black hole 
apparent horizon. This is done for a single non rotating black hole 
(Schwarzschild metric) in Section II. The interior metric in this case is the
spatial part of a positive curvature Friedmann-Robertson-Walker (FRW) metric,
so that the resulting construction can be interpreted as initial data for the
Oppenheimer-Schneider dust collapse (Section III). In Section IV, we give in 
closed form the corresponding solution for two non-rotating stuffed black 
holes (the analogous of the Misner data). We provide in Section V that a step 
by step procedure to stuff time symmetric initial data containing any number 
of black holes with arbitrary sizes and locations.

Finally, in the Appendix, we compare the numerical evolution of a 'plain' black
hole with that of a 'stuffed' black hole in the spherically symmetric case.
This is a biased test, because spherical coordinates are singular at the
origin and this coordinate singularity affects only the stuffed hole, because
in that case there is no excised region and the computational domain includes
the origin. In spite of that, we have evolved stuffed holes with the same
accuracy and stability as plain holes. This opens the door to 
three-dimensional applications, where the 'stuffing' approach
avoids putting internal curvilinear boundaries that would need a special
treatment. The advantage of the 'stuffed' black holes versus the 'plain' ones
will be manifest if the coordinate system contains a shift vector which allows
the holes to move across the numerical grid: in the plain case one would need
to treat a number of moving curvilinear boundaries, but in the stuffed case
nothing special is to be done because there are no internal boundaries  in that
case.

\section{SINGLE HOLE INITIAL DATA}

The evolution formalism~\cite{Lich,Choquet,ADM} is specially well suited for 
Numerical Relativity. In normal coordinates, the spacetime metric can be 
written as
\begin{equation}
ds^2 = -\alpha^2\;dt^2 + \gamma_{ij}\;dx^i\;dx^j
\;.	\label{metric}
\end{equation}
To start the evolution one needs both the spatial metric $\gamma_{ij}$ on the 
initial hypersurface and its second fundamental form (extrinsic curvature)
\begin{equation}
K_{ij} = -{1 \over 2\alpha}\;\partial_t \gamma_{ij} \;.
\label{Kij}
\end{equation}

These initial data, however, are constrained by the following equations
\begin{equation}
16\pi\;\tau = 16\pi\;\alpha^2 \;T^{00} = 
		R^j_{\;j} + (K^j_{\;j})^2 - K^j_{\;k}K^k_{\;j}
\label{energy_constraint}         
\end{equation}
(energy constraint), and
\begin{equation}
8\pi\;S_i = 8\pi\;\alpha \; T^0_{\;i} = K^j_{\;i;j} - \partial_i(K^j_{\;j})
\label{momentum_constraint}
\end{equation}
(momentum constraint), where both the Ricci tensor $R_{ij}$ and the covariant
derivatives are the ones corresponding to the three-dimensional geometry of the
initial slice. These constraint equations are first integrals of the Einstein
equations: they are verified at every time slice by the spatial part of any
spacetime metric.

One can learn a lot about the spacetime by just looking at the geometry of the
spatial hypersurfaces. Let us consider for instance a closed two-surface $\sigma$
in one such a three-dimensional manifold. The expansion $\theta$ of a
congruence of outgoing light rays starting at $\sigma$ is given by~\cite{York89}
\begin{equation}
\theta = n^j_{\;;j} + K_{ij}\,n^i\,n^j - K^j_{\;j}
\label{light_expansion}
\end{equation}
where the three-dimensional vector $n^i$ is the unit normal to $\sigma$. In the
'time symmetric' case ($K_{ij} = 0$) the minimal surfaces of the three-geometry
\begin{equation}
n^j_{\;;j} = 0
\label{minimal}
\end{equation}
are also apparent horizons ($\theta = 0$) and vice versa.

In the case of non-rotating black holes, one usually starts with time
symmetric initial data (and therefore zero initial momentum density $S_i$) and 
conformally flat initial metrics, namely
\begin{equation}
\gamma_{ij} = \psi^4\;\delta_{ij} \;.
\label{conformal}
\end{equation}
This simplifies the constraint equations~\cite{York72}, so that 
(\ref{energy_constraint}) can be written in terms of a flat space Laplace 
operator $\Delta$
\begin{equation}
\Delta\;\psi = -2\pi\; \tau\;\psi^5 \;.
\label{Laplace}
\end{equation}

In the spherically symmetric case, the regular vacuum solution of
(\ref{Laplace}) can be written as
\begin{equation}
\psi_{SW} = 1 + {m \over 2r} \;,
\label{single}
\end{equation}
which corresponds to the space part of the Schwarzschild metric in isotropic
coordinates. Allowing for (\ref{conformal}), the minimal surface (apparent 
horizon) condition (\ref{minimal}) reads
\begin{equation}
\partial_r (r\;\psi^2) = 0 \;,
\label{throat}
\end{equation}
which holds for $r_0 = m/2$. Note that the combination $r\;\psi^2$ is precisely
the 'area radius' $R$ of the spherical geometry, so that the apparent
horizon (minimal surface) is placed at
\begin{equation}
R(r_0) = 2m \;.
\label{horizon}
\end{equation}

Condition (\ref{throat}) can also be obtained by realizing that the metric 
given by (\ref{conformal},\ref{single}) is invariant under the discrete symmetry
\begin{equation}
r \;\;\;\; \longleftrightarrow \;\;\;\; r_0^2/r \;.
\label{inversion}
\end{equation}
which can be interpreted as a coordinate inversion at $r = r_0$. This inversion
symmetry provides suitable boundary conditions when one excises the spherical
region $r < r_0$ from the computational domain. The resulting inner boundary
condition is very easy to implement numerically in spherical coordinates, but
it is much more difficult to manage in the three-dimensional case (where one
would get a curvilinear boundary in cartesian coordinates) even for a single 
black hole.

As an alternative to this inner boundary aproach, we will construct a 'stuffed'
black hole by replacing the initial data (\ref{single}) with
\begin{equation}
\psi = \cases{
 1 + {\displaystyle m \over \displaystyle 2r} & for $ r>m/2 $ \cr
\mathstrut
\sqrt{\displaystyle 8 \over 
\displaystyle {1 + (2r/m)^2}} & for $ r<m/2 $ \cr}
\label{stuff}
\end{equation}
so that in the exterior region one recovers the 'plain' black hole solution 
(\ref{single}), whereas in the interior region one gets, allowing for
(\ref{conformal}), a homogeneous three-dimensional metric which is
the space part of a closed Friedman-Robertson-Walker (FRW) model (see
Fig. 1).

\section{PHYSICAL INTERPRETATION}

Up to now, we have just constructed time symmetric data on the initial slice.
This means that we have used only the energy and momentum constraints, but not
the remaining (evolution) field equations. From the energy constraint
(\ref{Laplace}), one can compute the energy density
\begin{equation}
\tau = \cases{
 0 & for $ r>m/2 $ \cr
\mathstrut
{\displaystyle 3 \over \displaystyle 32\pi\,m^2} & for $ r<m/2 $ \cr}
\label{tau}
\end{equation} 
whereas the momentum constraint (\ref{momentum_constraint}) plus the
time-symmetry condition ($K_{ij} = 0$) imply that the momentum density should
vanish. However, in order to get a physical interpretation of the matter
content of the spacetimes generated by these initial data, we need to say
something about the remaining (space) components of the stress-energy tensor
$T^{\mu\nu}$.

To do this, we can study the time evolution of the matching conditions between
the constant density interior and the vacuum exterior, namely
\begin{equation}
T^{\mu\nu}\;\Phi_\nu|_{\Phi=0} = 0 \;,
\label{Hugoniot}
\end{equation} 
where $\Phi = 0$ is the equation of the matching hypersurface. Allowing for the
fact that the momentum density vanishes, one gets easily from (\ref{Hugoniot})
that $\Phi$ can not depend on time. This means that the matching surface can be
taken to be the sphere
\begin{equation}
\Phi = r\,-\,m/2
\label{sphere}
\end{equation}
for any value of time. It follows then from (\ref{Hugoniot}) that the radial
direction gives an eigenvector of the stress-energy tensor (computed at the
matching surface) with zero eigenvalue.

If we assume for simplicity a barotropic perfect fluid matter content, 
it follows that the pressure should vanish (incoherent matter). 
The initial data (\ref{stuff}) can
then be understood as corresponding to a particular case of the well known 
Oppenheimer-Schneider dust collapse: a constant density spherical star which 
is initially at rest. In our case, the initial star radius actually coincides 
with the position of the apparent horizon.

The former is just one of the many possibilities allowed by the no-hair 
theorems. 
 A very interesting case is the 'string
perfect dust' matter content~\cite{Stachel}
\begin{equation}
T^{\mu\nu} = S^{\mu\rho}\;S_\rho^{\;\nu}\;,
\label{stringy}
\end{equation}
where 
\begin{equation}
S^{\mu\nu} = a^\mu\;b^\nu - b^\mu\;a^\nu
\label{bivector}
\end{equation}
is a simple surface-forming bivector.

There are basically two different possibilities to get a zero radial eigenvalue 
in the string case~\cite{Stachel}: either $S^{\mu r} = 0$ (the bivector has no
radial component) or the vector $\lambda^\mu = S^{\mu r}$ is isotropic (and,
being antisymmetric, has no radial component). In both ways, it follows that 
the string stress tensor breaks the spherical symmetry of the initial data, 
and this will lead to a non-spherical time evolution. This 'dynamical symmetry
breaking' is in contrast with the more predictable behaviour that one gets in
the dust case, where the full stress-energy tensor is spherically symmetric, so 
that the  spherical symmetry of the initial data will be preserved during 
evolution. 

\section{TWO BLACK HOLES INITIAL DATA}

The Misner initial data~\cite{Misner} are axially symmetric and describe two 
identical non-rotating black holes which are initially at rest. They can be
obtained by linear superposition of spherically symmetric solutions of the
Laplace equation, with centers distributed along the symmetry axis:
\begin{equation}
\psi = 1 + \sum_{n=1}^\infty ( \Phi^+_n + \Phi^-_n )
\label{sum}
\end{equation}
with
\begin{equation}
\Phi^\pm_n = {a \over r^\pm_n}\;\hbox{csch}(n\mu_0) \;,
\label{Phi}
\end{equation}
where 
\begin{equation}
(r^\pm_n)^2 = x^2 + y^2 + [ z \pm a \;\coth (n\mu_0) ]^2\;.
\label{radius}
\end{equation}

In order to see how this solution is obtained, let us notice that in the time
symmetric vacuum case the constraint (\ref{Laplace}) on the conformal factor
$\psi$ is the flat space Laplace equation. We know from electrostatics that we
can take advantage of the invariance of the Laplace equation under discrete
symmetries, such as (\ref{inversion}), by using the 'imaging method'~\cite{Smythe}.
This technique was adapted to Relativity by Misner~\cite{Misner2} in order to 
obtain initial data which are invariant by inversion across a 
number of spherical surfaces, which will become at the end minimal surfaces
(apparent horizons) of the resulting solution. This solution can be 
interpreted as describing time symmetric 
initial data for a number of black holes.

The apparent horizons (minimal surfaces) in (\ref{sum}) are the two spheres 
$\sigma^\pm$ given by 
\begin{equation}
r^\pm_1 =  a\;\hbox{csch}(\mu_0)\;.
\label{2horizons}
\end{equation}
The term $\Phi^+_{n+1}$ has been constructed as the image of $\Phi^-_n$ 
under inversion across $\sigma^+$. Also, the term $\Phi^-_{n+1}$ has been 
constructed as the image of $\Phi^+_n$ under inversion across $\sigma^-$. 
This means that the dipole combinations
\begin{equation}
\Lambda^\pm_n = \Phi^\pm_n + \Phi^\mp_{n-1} 
\label{dipole}
\end{equation}
are invariant under inversion across the sphere $\sigma^\pm$, respectively.

We have seen in the previous section how to stuff a single black hole. 
Now we have two identical holes and we will look for regular interior solutions
for each one. Let us begin by considering the interior region to the first
sphere $\sigma^-$. The infinite sum in the Misner
solution (\ref{sum}) can here be written in a more convenient form
\begin{equation}
\psi = \sum_{n=1}^\infty \Lambda^-_n
\label{newsum}
\end{equation}
so that every term is now invariant under inversion across the sphere 
$\sigma^-$. We will make use of this symmetry property to match every term in 
(\ref{newsum}) to (the conformal factor of) a constant curvature metric.

The conformal factor for a constant (positive) curvature metric
(closed FRW) can be written as
\begin{equation}
\psi_{FRW} = \sqrt{\displaystyle F\;\lambda \over 
\displaystyle {1 - 2\,b\,r^-_1\,\cos \varphi + (b^2+\lambda^2/4) (r^-_1)^2}}\;,
\label{FRW}
\end{equation}                                        
where $F$ is an arbitrary scale factor and $\lambda$ and $b$ are parameters
related to the conformal transformations of Euclidean three-space. If we impose
inversion symmetry (Eq.~\ref{throat}) across the sphere $\sigma^-$, we get
\begin{equation}
(b^2+\lambda^2/4)\;a^2 = \sinh^2\mu_0\;.
\label{1matching}
\end{equation}
This single condition ensures, allowing for (\ref{throat}), that the normal 
derivatives of the vacuum and FRW conformal factors will coincide on 
$\sigma^-$ if and only if both conformal factors actually coincide there.

It follows that, in order to complete the matching, we need only to look at 
the conformal factor values at $\sigma^-$ and tune the free parameters in 
(\ref{FRW}). The FRW conformal factor can be easily evaluated on $\sigma^-$:
\begin{equation}
\psi_{FRW}|_{\sigma^-} = \sqrt{\displaystyle F\;\lambda/2 \over 
\displaystyle {1 - b\,(z-a\,\coth\mu_0)} }\;, 
\end{equation}
where we have noted
\begin{equation}
z-a\,\coth\mu_0 = r^-_1\,\cos \varphi\;. 
\end{equation}
On the other hand, a straightforward calculation shows that
\begin{equation}
\Lambda^-_n|_{\sigma^-} = 2\,\Phi^-_n|_{\sigma^-} = 
{\displaystyle 2\,a \over \displaystyle 
\sqrt{a^2 + 2a\,z\,\sinh[n\mu_0]\sinh[(n-1)\mu_0]/\sinh[\mu_0]} }\;.
\end{equation}

Now it is easy to obtain the values of the arbitrary parameters in (\ref{FRW})
that ensure the matching between the vacuum dipole term $\Lambda^-_n$ 
and the conformal factor
\begin{equation}
\psi^-_n = \sqrt{\displaystyle F_n\;\lambda_n \over 
\displaystyle {1 - 2\,b_n\,(z-a\,\coth\mu_0) + \sinh^2\mu_0\,(r^-_1/a)^2}}
\label{FRW-}
\end{equation}                                        
across $\sigma^-$ for every value of $n$. 
Their actual values are:
\begin{eqnarray}
a\,b_n &=& -\sinh \mu_0 \;{ \displaystyle {\cosh[(2n-1)\mu_0]-\cosh[\mu_0]} 
\over \displaystyle {\cosh[(2n-1)\mu_0]\;\cosh[\mu_0] - 1} }  \\
a\,\lambda_n &=&  
{ \displaystyle {2\;\sinh[(2n-1)\mu_0]\;\sinh^2[\mu_0]}\over 
\displaystyle {\cosh[(2n-1)\mu_0]\;\cosh[\mu_0] - 1} }  \\
F_n &=& { \displaystyle {4\;a}\over 
\displaystyle {\sinh[(2n-1)\mu_0]} }.
\end{eqnarray}

Allowing for the symmetry of the solution across the equatorial plane, it is
easy to see that the conformal factor
\begin{equation}
\psi^+_n = \sqrt{\displaystyle F_n\;\lambda_n \over 
\displaystyle {1 + 2\,b_n\,(z+a\,\coth\mu_0) + \sinh^2\mu_0\,(r^+_1/a)^2}}
\label{FRW+}
\end{equation}                                        
will also match the vacuum dipole term $\Lambda^+_n$ across $\sigma^+$.
It follows that the complete solution for the stuffed two hole problem reads 
\begin{equation}
\Psi  = \cases{
\sum_{n=1}^\infty \psi^-_n & for $ r^-_1 < a\;\hbox{csch}(\mu_0) $ \cr
\sum_{n=1}^\infty \psi^+_n & for $ r^+_1 < a\;\hbox{csch}(\mu_0) $ \cr
1 + \sum_{n=1}^\infty ( \Phi^+_n + \Phi^-_n ) & elsewhere. \cr}  
\label{2stuff}
\end{equation}

The resulting solution is regular and smooth everywhere (see Fig.~2). Following
Bowen {\it et al}\cite{Bowen}, we can define an effective energy density
starting from a 'Newtonian potential' linearly related to $\Psi$. In our case, 
allowing for (\ref{Laplace}), this amounts to
\begin{equation}
\tau_{eff} = \tau \; \Psi^5  \;.
\label{taueff}
\end{equation} 
This effective energy density can be easily expressed as the sum of the
effective energy density of every FRW component, namely
\begin{equation}
\tau_{eff} = \cases{ \sum_{n=1}^\infty (\tau_{eff})_n 
& for $ r^{\pm}_1 < a\;\hbox{csch}(\mu_0) $ \cr
 0 & elsewhere \cr} 
\label{2Dtau}
\end{equation} 
where
\begin{equation}
(\tau_{eff})_n = {\displaystyle 3 \over \displaystyle 8\pi\,F_n^2} \;
\left( \psi^{\pm}_n \right) ^5 \; .
\label{FRWtau}
\end{equation}
The positivity of the factors $\psi_n^{\pm}$ ensures that the energy density
(\ref{2Dtau}) is positive inside the holes and it is bounded by the maximum 
energy density of the FRW components.

As in the single hole case, the energy density has a jump at the matching 
surfaces due to the discontinuity of the second metric derivatives there.
It follows from the previous section that a reasonable physical description of 
the solution could be that of two balls of incoherent matter. 
Notice, however, that the energy
density of every ball is not constant, as it was in the single hole case. The
gravitational interaction between the two balls accounts for their distortion.

\section{MULTIPLE BLACK HOLE CASE}

Let us consider now the time symmetric multiple black hole case. As is well
known, the vacuum exterior solution can be obtained by the conformal imaging
method~\cite{Misner2}. 
As an input for this method, one must provide the size and 
location of an arbitrary number $N$ of spheres, which will become at the end 
the apparent horizons of $N$ black holes. The resulting solution, by 
construction, will be then inversion symmetric across the apparent horizon of 
every black hole. It could be written as a linear superposition of poles,
\begin{equation}
\Psi  = 1 + \sum_{n=1}^\infty { \displaystyle a_n 
                          \over \displaystyle |\vec x - \vec x_n| }\;.
\label{multipole}
\end{equation}

Our goal is to provide a suitable interior solution for every black hole. As
far as the holes do not overlap, we can consider them separately. Let us begin
with the first one: its horizon $\sigma^{(1)}$ is a sphere centered at $\vec x_1$,
\begin{equation}
 |\vec x - \vec x_1| = R_1  \;.
\label{3Dhorizon}
\end{equation}
Note that, as (\ref{multipole}) is invariant under inversion across the sphere
$\sigma^{(1)}$, half of the poles in (\ref{multipole}) are outside 
and half inside. Moreover, every outside pole is the image under inversion of 
an inside pole and vice versa. 

This means that we can combine every pole with its image to form the invariant 
dipoles 
\begin{equation}
\Lambda_j^{(1)} =  { \displaystyle a_j \over \displaystyle |\vec x - \vec x_j| }
     +  { \displaystyle a_{j'} \over \displaystyle |\vec x - \vec x_{j'}| }\;,
\label{3Ddipole}
\end{equation}
where the exterior point $x_{j'}$ is the image of the interior one $x_j$ under
inversion across the sphere $\sigma^{(1)}$. The sum in the solution 
(\ref{multipole}) can then be rearranged as follows
\begin{equation}
\Psi  = \sum_{j=1}^\infty \Lambda_j^{(1)} \;.
\label{multidipole}
\end{equation}

Our strategy will be then to match separately every invariant dipole
(\ref{3Ddipole}) to a closed FRW factor, given by
\begin{equation}
\psi^{(1)}_j = \sqrt{\displaystyle F_j\;\lambda_j \over 
\displaystyle {1 - 2\,\vec b_j\,(\vec x-\vec x_1) 
 + (b_j^2+\lambda_j^2/4) |\vec x-\vec x_1|^2}}\;,
\label{3DFRW}
\end{equation}                                        
where
\begin{equation}
\vec b_j = b_j\;{\displaystyle  \vec x_j - \vec x_1 \over 
             \displaystyle |\vec x_j - \vec x_1|} \;.
\end{equation}                                        

As in the previous section, we must impose first that the interior solution
(\ref{3DFRW}) have a 'throat' (\ref{throat}) at the sphere $\sigma^{(1)}$, 
obtaining the condition
\begin{equation}
(b_j^2+\lambda_j^2/4) = 1/R_1^2\;,
\label{matching3D}
\end{equation}                
which generalizes (\ref{1matching}). As in the previous section, this condition
ensures that the radial derivatives of the vacuum and the FRW conformal factors
will coincide on the sphere $\sigma^{(1)}$ if and only if both conformal 
factors actually coincide there.

Alowing for (\ref{matching3D}), the interior factor (\ref{3DFRW}) at
the horizon is
\begin{equation}
\psi^{(1)}_j|_{\sigma^{(1)}} = \sqrt{\displaystyle F_j\;\lambda_j/2 \over 
\displaystyle {1 - \vec b_j\,(\vec x - \vec x_1) } }\;.
\end{equation}                              
On the other hand, allowing for the fact that the two images in the invariant 
dipole (\ref{3Ddipole}) coincide by construction on the sphere $\sigma^{(1)}$, 
we have
\begin{equation}
\Lambda_j^{(1)}|_{\sigma^{(1)}} =  { \displaystyle 2a_j \over 
\sqrt{ \displaystyle { R_1^2 + 2 (\vec x - \vec x_1)(\vec x_1 - \vec x_j) +
 |\vec x_1 - \vec x_j|^2 } } }\;.
\end{equation}

It follows that the choice of parameters
\begin{eqnarray}
b_j &=& -2{ \displaystyle |\vec x_1 - \vec x_j| \over 
\displaystyle {R_1^2  + |\vec x_1 - \vec x_j|^2 } } \\
\lambda_j &=&  2/R_1\;{ \displaystyle {R_1^2  - |\vec x_1 - \vec x_j|^2 } \over 
\displaystyle {R_1^2  + |\vec x_1 - \vec x_j|^2 } } \\
F_j\,\lambda_j &=& { \displaystyle {8\;a_j}\over 
\displaystyle {R_1^2  + |\vec x_1 - \vec x_j|^2 } }  
\end{eqnarray}
ensures the matching between (\ref{3Ddipole}) and (\ref{3DFRW}) at 
$\sigma^{(1)}$.

The interior solution for the first hole can then be obtained by summing the
corresponding FRW factors (\ref{3DFRW})
\begin{equation}
\Psi^{(1)}  = \sum_{j=1}^\infty \psi_j^{(1)} \;.
\end{equation}
The interior solution for every other hole can be obtained in exactly the same 
way.

\appendix
\section*{1D Numerical Evolution}       
We will compare here the numerical evolution of a single spherically symmetric
(1D) stuffed black hole against a 'plain' one. We use in both cases the same
finite difference code with a $400$ point evenly space numerical grid. The
metric quantities and their evolution equations are also the same. The matter
density and speed are computed in the stuffed case by using the standard
'upwind' method to model the continuity equation and the Euler equation for
dust, respectively (they are equivalent to the stress-energy tensor 
conservation).

In the 'plain' case, we have used the throat inversion symmetry to provide the
inner boundary condition. In the general (3D) stuffed case there is no inner
boundary. The only concern will arise from the discontinuity of the energy
density (second derivatives of the metric) at the throat. We have not seen any
problem with the matter variables (density and speed) in our 1D numerical
evolution.

The singularity of spherical coordinates at $r=0$, however, demands a
special treatment of the origin in the 1D stuffed case. We have done it by
adding two virtual points as the mirror image of the first two grid points 
across the origin. In the 'plain' hole case the origin is contained in the
excised region and such special treatment is not needed. This is why this 1D
test is biased: it is the best case for the plain hole (inversion symmetry in
spherical coordinates) and the worst one for the stuffed case (need for a
special treatment of the origin).

We have compared the code performance, in terms of accuracy and stability, in 
both cases for three different slicing conditions: harmonic, '1+log' and
maximal. In the last two cases, we can barely notice any significant difference
between the plain and the stuffed hole evolution in the exterior region (see
Fig.~3). In the harmonic slicing case, however, the code for the stuffed black 
hole crashes as the dust ball in the interior region collapses, evolving 
towards a singularity.
This is a consequence of the weaker singularity avoidance properties of 
harmonic slicing.

These results are promising with regards to 3D applications in rectilinear
grids, where inversion symmetry across the spherical horizons is not so easy
to implement (plain case), whereas there is no internal boundary of any kind
(the coordinate system is regular everywhere) in the stuffed case.


\acknowledgments
 This work is supported by the Direcci\'on General para 
Investigaci\'on Cient\'{\i}fica y T\'ecnica of Spain under project PB94-1177

\begin{figure}
\caption{Plot of the conformal factor describing initial data for a single
stuffed black hole with mass $m = 2$ (apparent horizon at $r = 1$ in 
isotropic coordinates). The smoothness and regularity of the solution is 
evident.}
\end{figure}

\begin{figure}
\caption{Surface plot of the conformal factor describing initial data for 
two stuffed black holes with $\mu_0 = 2$, $a=1$. The plot is in cylindrical
coordinates with the axis along the line joining the two holes. The azimuthal 
angle $\phi$ is suppressed, as the solution is axially symmetric. The 
smoothness and regularity of the solution is evident.}
\end{figure}

\begin{figure}
\caption{Evolution of plain (solid line) and stuffed (dashed line) black holes 
in the spherically symmetric case ('1+log' slicing, $400$ grid points). 
Left: the radial metric coefficient is plotted after $t=300m$. 
Right: the maximum error for the black
hole mass near the horizon is plotted as a function of time. Both the stuffed 
and the plain case show the same accuracy.}
\end{figure}


\begin{references}
\bibitem{Misner}	C.W.~Misner, (1960):
               		Phys.\ Rev. {\bf 118}, 1111.
\bibitem{Smarr}		L.~Smarr, A.~Cadez, B.~DeWitt and K.~Eppley, (1976):
               		Phys.\ Rev.\ D {\bf 14}, 2443.
\bibitem{Lindquist}	R.W.~Lindquist, (1963):
			J.\ Math.\ Phys. {\bf 4}, 938.
\bibitem{Lich}		A.~Lichnerowicz, (1944):
	 		J. Math. Pures Appl. {\bf 23}, 37.
\bibitem{Choquet}	Y.~Choquet-Bruhat, (1962): in 
	 		{\it Gravitation: an introduction to Current Research},
	 		ed. L.~Witten (Wiley, New York).   
\bibitem{ADM}		R.~Arnowitt, S.~Deser and C.~W.~Misner, (1962): in 
	 		{\it Gravitation: an introduction to Current Research},
	 		ed. L.~Witten (Wiley, New York).   
\bibitem{York89}	J.~W.~York Jr., (1989): in
                        {\it Frontiers in Numerical Relativity},
			ed. C.~Evans, L.~Finn and D.~Hobill (Cambridge U.P.).
\bibitem{York72}	J.~W.~York Jr., (1972):
	 		Phys. Rev. Lett. {\bf 28}, 1082.
\bibitem{Stachel}	J.~Stachel, (1980):
	 		Phys.\ Rev.\ D {\bf 21}, 2171.
\bibitem{Smythe}	W.R.~Smythe, (1950):
			{\it Static and Dynamic Electricity} 
			(Mac Graw-Hill, New York).
\bibitem{Bowen}		J.~Bowen, J.D.~Rauber, and J.W.~York, (1984):
			Class. Quantum Grav. {1}, 591.
\bibitem{Misner2}	C.W.~Misner, (1963):
                        Ann. Phys. (N.Y.) {\bf 24}, 102.

\end{references}
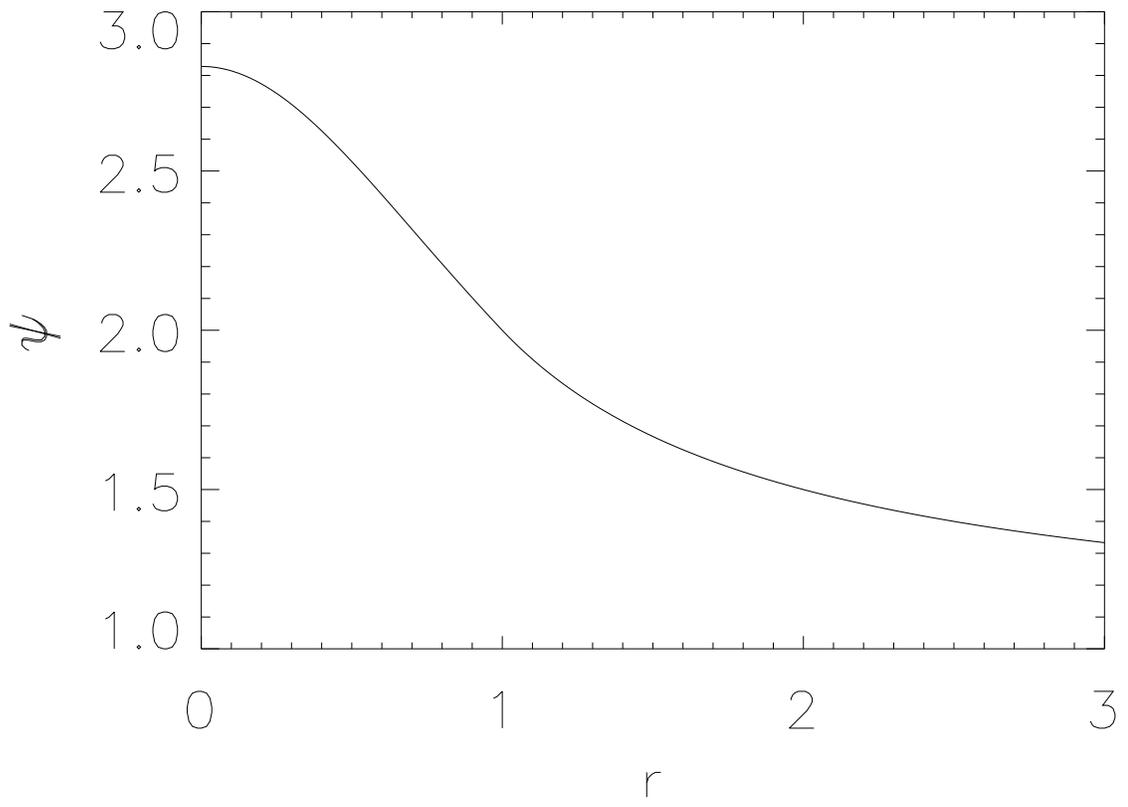
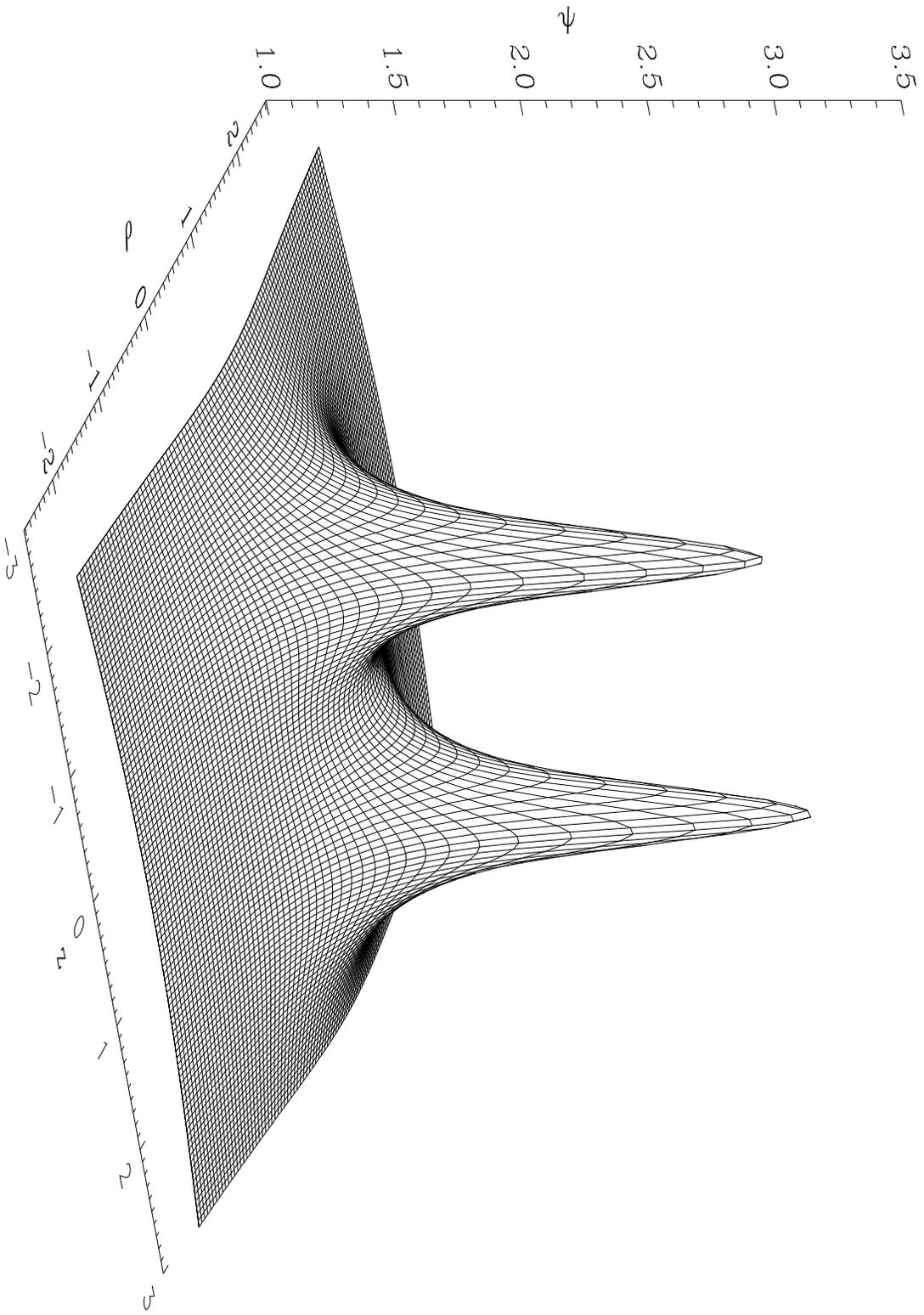
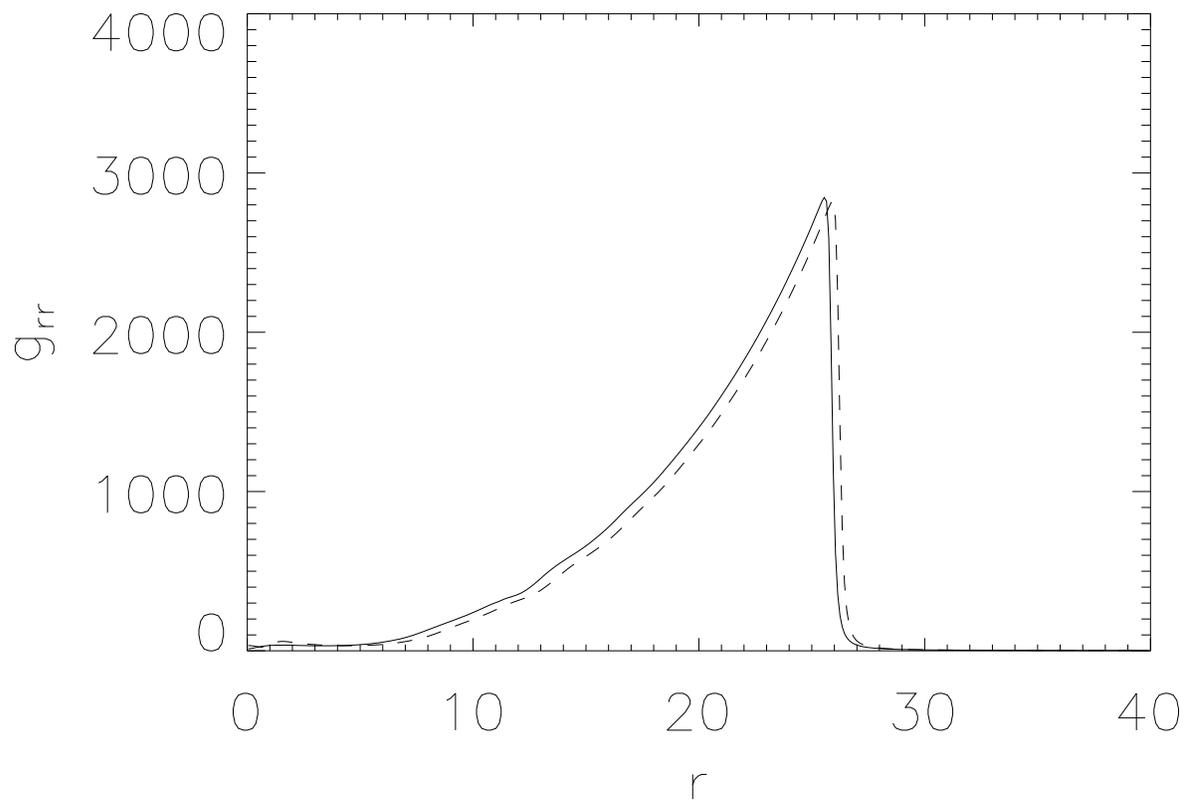
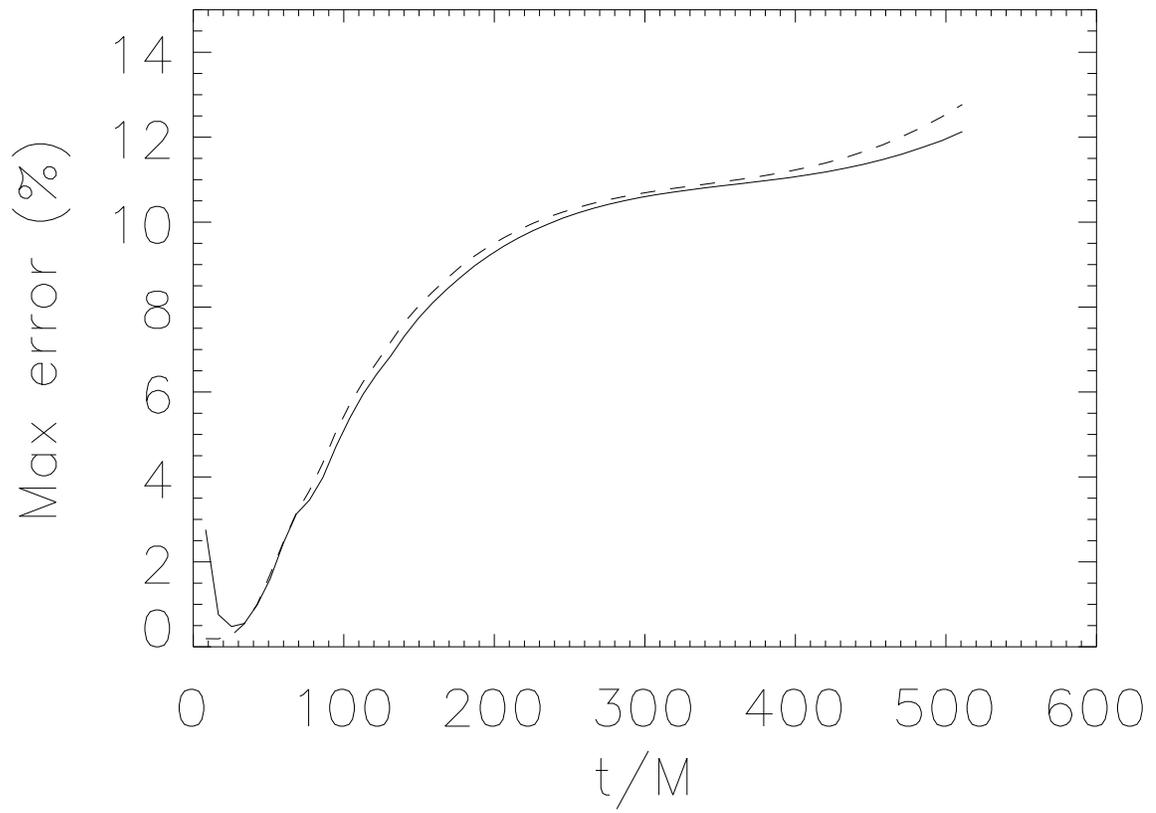
\end{document}